\documentclass[sigconf,nonacm]{acmart}

\usepackage{balance} % for creating a balanced last page

\usepackage{hyperref}
\usepackage{graphicx}
\usepackage{xspace}
\usepackage{amsmath}  % needed for \frac ?
\newcommand{\fullonly}[1]{}

\newcommand{\ieeeonly}[1]{}
\newcommand{\lncsonly}[1]{}
\newcommand{\articleonly}[1]{}
\newcommand{\acmonly}[1]{#1}
\newcommand{\svonly}[1]{}

\newcommand{\mycomment}[1]{}

\articleonly{
\usepackage{setspace}
\setstretch{1.1}
\usepackage{fullpage} %or \usepackage[margin=1.0in]{geometry}
}

\articleonly{}
\lncsonly{}
\acmonly{}

\begin{document}

\acmonly{
\copyrightyear{2025}
\acmYear{2025}
\setcopyright{acmlicensed}\acmConference[SACMAT '25]{Proceedings of the 30th ACM Symposium on Access Control Models and Technologies}{July 8--10, 2025}{Stony Brook, NY, USA}
\acmBooktitle{Proceedings of the 30th ACM Symposium on Access Control Models and Technologies (SACMAT '25), July 8--10, 2025, Stony Brook, NY, USA}
\acmDOI{10.1145/3734436.3734441}
\acmISBN{979-8-4007-1503-7/2025/07}
}

% following is copied from spaa-19-sig.tex, the template recommended at http://www.scomminc.com/pp/acmsig/sacmat.htm
% it needs to appear after the above \acm* commands
\fancyhead{}

\title{ABAC Lab: An Interactive Platform for Attribute-based Access Control Policy Analysis, Tools, and Datasets
\ieeeonly{\thanks{\thanksText}}\lncsonly{\thanks{\thanksText}}\articleonly{\thanks{\thanksText}}
}

\subtitle{[Dataset/Tool Paper]}
% do not use \titlenote when using \thanks
%\acmonly{\titlenote{Produces the permission block, and copyright information}}

\acmonly{
\author{Thang Bui}
\affiliation{
\institution{Cal State Monterey Bay}
\city{Seaside}
\state{CA}
\country{U.S.A.}}
\email{cbui@csumb.edu}
\author{Anthony Matricia}
\affiliation{
\institution{Cal State Monterey Bay}
\city{Seaside}
\state{CA}
\country{U.S.A.}}
\email{amatricia@csumb.edu}
\author{Emily Contreras}
\affiliation{
\institution{Cal State Monterey Bay}
\city{Seaside}
\state{CA}
\country{U.S.A.}}
\email{emicontreras@csumb.edu}
\author{Ryan Mauvais}
\affiliation{
\institution{Cal State Monterey Bay}
\city{Seaside}
\state{CA}
\country{U.S.A.}}
\email{rmauvais@csumb.edu}
\author{Luis Medina}
\affiliation{
\institution{Cal State Monterey Bay}
\city{Seaside}
\state{CA}
\country{U.S.A.}}
\email{luismedina4@csumb.edu}
\author{Israel Serrano}
\affiliation{
\institution{Cal State Monterey Bay}
\city{Seaside}
\state{CA}
\country{U.S.A.}}
\email{iserrano@csumb.edu}
}

% ACM sigconf format also allows the following info for each author:
%\department{Department of Computer Science}
%\streetaddress{New Computer Science, MSC 2424}
%\city{Stony Brook}
%\state{NY}
%\postcode{11794-2424}
%\country{U.S.A.}

% acm: abstract must precede maketitle
% lncs (and probably other styles): abstract must follow maketitle.
\newcommand{\abstracttext}{ %
Attribute-Based Access Control (ABAC) provides expressiveness and flexibility, making it a compelling model for enforcing fine-grained access control policies. To facilitate the transition to ABAC, extensive research has been conducted to develop methodologies, frameworks, and tools that assist policy administrators in adapting the model. Despite these efforts, challenges remain in the availability and benchmarking of ABAC datasets. Specifically, there is a lack of clarity on how datasets can be systematically acquired, no standardized benchmarking practices to evaluate existing methodologies and their effectiveness, and limited access to real-world datasets suitable for policy analysis and testing.

This paper introduces ABAC Lab, an interactive platform that addresses these challenges by integrating existing ABAC policy datasets with analytical tools for policy evaluation. Additionally, we present two new ABAC datasets derived from real-world case studies. ABAC Lab serves as a valuable resource for both researchers studying ABAC policies and policy administrators seeking to adopt ABAC within their organizations. By offering an environment for dataset exploration and policy analysis, ABAC Lab facilitates research, aids policy administrators in transitioning to ABAC, and promotes a more structured approach to ABAC policy evaluation and development.
}

\acmonly{
\begin{abstract}
\abstracttext
\end{abstract}}

\ccsdesc[500]{Security and privacy~Access control}

\acmonly{\keywords{policy analysis tool; attribute-based access control; dataset}}

\acmonly{\thanks {This material is based on work supported in part by FY24-25 CSU RSCA CSUMB and COS DCI Grants. \\ © Thang Bui et al., 2025. This is the author's version of the work. It is posted here for personal use. Not for redistribution. The definitive version was published in SACMAT '25, \url{https://doi.org/10.1145/3734436.3734441}}}

% \acmonly{\thanks {\thanksText}}

\maketitle
\ieeeonly{
\begin{abstract}
\abstracttext
\end{abstract}}
\lncsonly{
\begin{abstract}
\abstracttext
\end{abstract}}
\articleonly{
\begin{abstract}
\abstracttext
\end{abstract}}
\svonly{
\begin{abstract}
\abstracttext
\end{abstract}}

% !TeX root = main.tex

\section{Introduction}
\label{sec:intro}

ABAC defines policies using attributes related to users, resources, actions, and environmental conditions. This flexibility and expressiveness make ABAC well-suited for complex and dynamic environments, where traditional access control models struggle to scale \cite{NIST14ABAC-URLshort}. Moreover, ABAC promises long-term cost savings by reducing the administrative burden associated with policy management. As a result, organizations are increasingly considering ABAC as a replacement for legacy access control mechanisms.

While ABAC offers significant advantages, research on its adoption and implementation often faces challenges related to dataset availability and policy analysis. The survey by \cite{parkinson2022survey} on empirical security analysis of access control systems identifies major obstacles in evaluating ABAC policies, such as difficulties in acquiring real-world datasets due to privacy and security concerns and the lack of standardized benchmarks for comparing policy analysis techniques.  Similarly, a survey on machine learning approaches in access control highlights that ML-based approaches for access control suffer from a lack of high-quality, publicly available datasets, making it difficult to train and evaluate models effectively \cite{nobi2022survey}. These limitations hinder the ability to assess different approaches effectively, slowing advancements in ABAC research and its broader adoption. 

To address these challenges, we introduce \textbf{ABAC Lab}, an open-source platform designed to support researchers and policy administrators working with ABAC policies. ABAC Lab offers a repository of ABAC datasets, including both existing and new real-world datasets, along with ABAC policy analysis tools and a publicly accessible GitHub repository. 

The dataset repository includes existing ABAC policy datasets from the literature, standardized into a unified format to facilitate benchmarking and comparison. It also allows researchers to contribute new datasets, fostering a collaborative research environment. In addition to integrating prior datasets, ABAC Lab introduces new datasets derived from real-world case studies. All datasets include detailed documentation on policy structures and attribute definitions. To assist with policy analysis, ABAC Lab provides tools for visualization, debugging, and testing, helping researchers explore ABAC policies and administrators manage policy transitions in real-world systems.

ABAC Lab is accessible through the GitHub repository: \url{https://github.com/ABAC-Lab-Admin/ABAC-Lab}. The repository includes detailed dataset descriptions, user guides, and a demo video showcasing the main features of ABAC Lab.

The main contributions of this paper are the introduction of ABAC Lab, an open-source platform for ABAC policy analysis, and the development of new real-world datasets for evaluating ABAC policies. In addition to these datasets, ABAC Lab includes a graphical user interface (GUI) and a suite of policy analysis features that, to the best of our knowledge, were not included in previously released tools or software associated with existing ABAC policy mining and analysis approaches (e.g., \cite{xu14miningABAClogs, xu15miningABAC}). 

% !TeX root = main.tex

\section{Background and Related Work}
\label{sec:related-work}
This section discusses related research in ABAC, with a focus on ABAC policy mining, which benefits most directly from ABAC Lab, along with the availability of ABAC datasets.

\subsection{Related work on ABAC policy mining}
\label{sec:abac-mining}

ABAC policy mining aims to automate the transition to ABAC by generating ABAC policies from existing lower-level access control data, such as access control lists (ACLs) or access logs. Instead of manually designing policies, which is costly and error-prone, policy mining algorithms extract concise, high-level ABAC rules that replicate existing access decisions. This approach significantly reduces the administrative burden of adopting ABAC.

To evaluate policy mining methods, researchers use benchmark datasets containing a full ABAC policy, consisting of user attributes, resource attributes, and a set of rules. From these rules, ground truth permissions for ACLs and log-based access decisions are generated, forming the input to the policy mining process. The mined ABAC policy rules produced by the algorithm are then compared against the ground truth rules. Importantly, these ground truth rules are not provided as part of the input for policy mining but serve as an evaluation metric to assess how accurately the mined policy reconstructs the original access control decisions.

There is substantial research in ABAC policy mining, as surveyed in \cite{das2018}. The first algorithm for ABAC policy mining was introduced by Xu et al. \cite{xu15miningABAC}, whose policy language and datasets we incorporate in ABAC Lab. We chose this policy language because it is more expressive than those used in other ABAC mining studies. Additionally, the datasets from \cite{xu15miningABAC} are publicly available and have been widely adopted in subsequent research. Using these datasets, Medvet et al. \cite{medvet2015} developed the first evolutionary algorithm for ABAC policy mining, while Iyer et al. \cite{iyer2018} introduced the first algorithm that can mine both permit and deny rules. These datasets have also been used in several other works \cite{karimi2022autoextract, jabal20learning, shang24mining, karimi18mining, talukdar2017efficient, das2019policy, mitani2023qualitative, bui24missingattr, bui2020learning}. This broad adoption further supports the role of these datasets as a common benchmark in the field.

Mining policies from access logs has also been explored. Xu et al. \cite{xu14miningABAClogs} proposed an algorithm leveraging the datasets from \cite{xu15miningABAC}. Law et al. \cite{law20fastlas} presented a scalable inductive logic programming algorithm for learning ABAC rules from logs and used one of the datasets from \cite{xu15miningABAC} for evaluation. Cotrini et al. \cite{sparselogs2018} introduced an approach using APRIORI-SD, a machine-learning technique for subgroup discovery. However, they did not use the datasets from \cite{xu14miningABAClogs} or \cite{xu15miningABAC}. Instead, they used a university policy dataset from their institution that is not publicly available. They explain that the reason for not using Xu et al.'s datasets is that the datasets contain a high proportion of access requests rather than sparse logs. To address this issue, ABAC Lab provides a log generation feature that allows researchers to create synthetic access logs under different configurations, including request distributions. 

Several work, including \cite{sparselogs2018} and \cite{law20fastlas}, have used Amazon datasets \cite{kaggleAmazoncomEmployee, uciAmazonDataset}, but we did not include them in our dataset repository due to their incompatibility with the ABAC policy language implemented in ABAC Lab. The dataset in \cite{kaggleAmazoncomEmployee} contains access decisions (permit or deny) for employee requests to various Amazon resources and was created for a Kaggle competition on building models to predict access decisions. However, it provides only eight anonymized user attributes and no resource attributes, with resources represented by IDs only. Additionally, the dataset lacks a rule set, which means there are no ground truth rules or permissions, which are essential for the evaluation process described earlier. Similarly, the dataset in \cite{uciAmazonDataset} is not ABAC in nature, as analyzed in \cite{nobi2022toward}. It lacks clear attribute data and a rule set, and there are cases where users with identical attributes have different access permissions. As further analyzed in \cite{cappelletti2019quality}, both datasets are not suitable for generating interpretable ABAC rules with reasonable accuracy.

\subsection{Expanding ABAC Datasets}
\label{sec:new-datasets-related-work}

To the best of our knowledge, the policies in \cite{xu15miningABAC} remain the only public ABAC policy datasets that attempt to capture real-world access control scenarios, while other studies have used different policies, mostly relying on synthetic datasets. However, the policies in \cite{xu15miningABAC} are manually crafted based on the authors' analysis and remain relatively small in scale. To expand the availability of more realistic ABAC datasets, ABAC Lab introduces two new datasets based on real-world case studies that provide large, more representative policy scenarios:

\begin{enumerate}
    \item E-Document \cite{decat14edoc} – A SaaS multi-tenant e-document processing platform with diverse access control requirements.
        
    \item Workforce Management \cite{decat14workforce} – A SaaS workforce management application that handles workflow planning and supply management for product or service appointments.
\end{enumerate}

\subsection{ABAC from Natural Language}
\label{sec:extract-abac-from-nl}

Another research direction involves extracting ABAC policies from natural language. Studies in \cite{alohaly2018extract, narouei2015extract} explore how machine learning and NLP techniques can automatically extract attributes and policies from textual documents. However, as noted in \cite{nobi2022survey}, the existing datasets used in this field are largely role-based and lack sufficient attribute diversity. To address this, researchers have expanded datasets by adding more subject- and object-related attributes, but they still fall short of fully representing ABAC policies.

In our own efforts, we applied ML and AI techniques from the literature to extract ABAC policies from the case studies mentioned in Section \ref{sec:new-datasets-related-work}.  Although the extracted policies contained some relevant elements, they lacked the structure needed for effective policy analysis, and our evaluation found them insufficiently accurate for practical use.
Due to these limitations, we do not currently include such datasets in ABAC Lab, but this will be explored in future work.

% !TeX root = main.tex

\section{Policy Language}
\label{sec:policy-language}

We adopt the datasets and the policy language introduced in \cite{xu15miningABAC}, as discussed in Section \ref{sec:abac-mining}, for ABAC Lab. In this section, we briefly describe the features of the policy language and refer the reader to \cite{xu15miningABAC} for details.

Policies in this language consist of rules defined over subject and resource attributes. It supports both single-valued and set-valued attributes, and includes operators such as equality, set membership, and set comparison. Rules can include constraints on subject attributes, resource attributes, and relationships between subject and resource attributes. The policy language offers a good balance between expressiveness and simplicity, making it suitable for modeling a wide range of realistic ABAC policies.

\section{ABAC Lab Datasets}
\label{sec:datasets}

ABAC Lab provides a curated collection of datasets to support ABAC policy mining and analysis. These datasets include publicly available datasets from prior research as well as new real-world case study datasets. Each dataset consists of user attributes, resource attributes, and a set of access control rules, structured in .abac files following the policy language described in Section \ref{sec:policy-language}. ABAC Lab is thoroughly documented, with detailed descriptions of the .abac format and policy structures for each dataset, including attribute definitions and rule explanations. All documentation is available in the platform repository. 

\subsection{Existing Public Datasets}
\label{sec: old-datasets}

ABAC Lab includes three ABAC policy datasets developed by Xu et al. \cite{xu15miningABAC}, which are widely used in ABAC policy mining research, as discussed in Section \ref{sec:abac-mining}.

\begin{itemize}
    \item \textit{University} – Manages access to university resources for students, instructors, teaching assistants (TAs), department chairs, and administrative staff. Resources include admission applications, gradebooks, transcripts, and course schedules.
    
    \item \textit{Project Management} – Manages access to project-related resources such as budgets, schedules, and tasks for department managers, project leaders, employees, contractors, auditors, accountants, and planners 
    
    \item \textit{Healthcare} – Manages access to electronic health records (EHRs) and individual record entries, with policies governing nurses, doctors, patients, and authorized agents (e.g., a patient’s spouse).
\end{itemize}

Table \ref{tab:policy-size} provides an overview of key statistics for these datasets, including the total number of rules, permissions, users, resources, and attributes.

\begin{table}[htb]
\begin{tabular}[b]{|l|l|l|l|l|l|l|}
\hline
\multicolumn{1}{|c|}{Policy} & \multicolumn{1}{c|}{\#sub} & \multicolumn{1}{c|}{\#res} & \multicolumn{1}{c|}{\#uAttr} & \multicolumn{1}{c|}{\#rAttr} & \multicolumn{1}{c|}{\#rule} & \multicolumn{1}{c|}{\#perm} \\ \hline
healthcare & 21 & 16 & 6 & 7 & 6 & 43 \\ \hline
project-mgmt & 19 & 40 & 8 & 6 & 5 & 101 \\ \hline
university & 22 & 34 & 6 & 5 & 10 & 168 \\ \hline
workforce & 353 & 250 & 10 & 16 & 28 & 15858 \\ \hline
e-document & 500 & 300 & 11 & 9 & 25 & 32961 \\ \hline
\end{tabular}
\caption{Policy sizes. \#sub, \#res, \#uAttr, \#rAttr, \#rule, \#perm are the number of users, the number of resources, the number of user attributes, the number of resource attributes, the number of rules in the policy, and the number of granted permissions by the policy, respectively.}
\vspace{-2em}
\label{tab:policy-size}
\end{table}

\subsection{New Real-World Case Study Datasets}
\label{sec: new-datasets}

As introduced in Section 2.2, ABAC Lab includes two new datasets based on real-world case studies designed to provide large-scale, realistic data for ABAC policy analysis and evaluation. In this section, we provide more details on these datasets and explain the process of developing them. 

Both datasets were translated from two large case studies developed by Decat et al., based on the access control requirements for Software-as-a-Service (SaaS) applications offered by real companies \cite{decat14edoc, decat14workforce}. The original case studies provided detailed natural-language descriptions of the policies, which we systematically translated into ABAC policies compatible with our framework. In addition to ABAC, these datasets were also used to derive Relationship-Based Access Control (ReBAC) policies and served as the basis for evaluating related work in ReBAC \cite{buirebacmining, iyerrebacmining} . Certain aspects, such as temporal conditions, obligations, and policy administration, were omitted and left for future work due to their complexity and lack of direct support in the current ABAC policy language.

The Workforce Management dataset captures a workforce management application, modeling organizational structures, workflows, and employee hierarchies. The policy defines roles such as technicians, managers, and operators and resources like work orders, tasks, and stock refill requests. Supported actions include creating work orders, completing tasks, and modifying resources based on operational needs.

The E-Document dataset models a secure document distribution application that handles both digital and printed documents for multiple organizations. The policy defines user roles such as employees, customers, and admins, interacting with document types like invoices, banking notes, and paychecks. It supports actions such as sending documents, receiving resources, and accessing meta-information.

Compared to the smaller, manually crafted datasets described in Section \ref{sec: old-datasets}, these new datasets offer a larger scale suitable for more robust evaluation, greater attribute diversity that better captures real-world complexity, and support for more complex attribute structures. As shown in Table \ref{tab:policy-size}, these new case studies significantly expand the size and complexity of ABAC Lab’s dataset collection.

% \textbf{\textit{Dataset Construction Methodology.}}

\paragraph{\textbf{Dataset Construction Methodology}} To build these datasets, we first identified the users and resources from the case studies and defined their corresponding attributes to ensure consistency and support policy rule construction. The extracted policies were then translated into the ABAC Lab policy language. 

After policy translation, we generated the object model, including users, resources, and their attribute values, using a policy-specific pseudorandom algorithm designed to produce realistic attribute data. This algorithm creates users and resources while selecting attribute values based on appropriate probability distributions that reflect realistic variability within each case study. The algorithm is parameterized by a set of size controls that determine key aspects of the dataset structure, such as the number of users, managers, tasks, or documents generated. These parameters allow flexibility in scaling the datasets for different experimental needs while preserving the logical relationships defined by the policies. Specific size parameters were configured for each dataset. For example, the Workforce Management dataset includes parameters controlling the number of managers, staff, work orders, and tasks, while the E-Document dataset uses parameters defining the number of employees, customers, and generated documents.

Ideally, using real attribute data from the organizations involved would further improve realism. However, obtaining such data is challenging, as organizations are generally unwilling to share internal information publicly, even in anonymized form, due to privacy and security concerns. Given this limitation, the object model generator was carefully designed based on the case study descriptions to produce realistic scenarios and attribute relationships that align with the original requirements.

The dataset generation code, including the object model generator, is available in the ABAC Lab repository. The repository also includes a generated dataset for each case study, reported in Table \ref{tab:policy-size}, and researchers can reuse the code to generate new datasets of varying sizes based on the same policies.

% !TeX root = main.tex
\section{ABAC Lab Features}
\label{sec:abac-lab}

ABAC Lab is designed to support both researchers and policy administrators in exploring and analyzing ABAC policies and datasets. Implemented in Python, the platform provides a desktop application that allows users to work directly with integrated datasets and perform various analytical tasks.

The features in ABAC Lab aim to simplify common challenges in ABAC research and policy management, such as understanding policy behavior, evaluating access decisions, and preparing data for experiments or review. Table~\ref{tab:abaclab-features} summarizes the main features of ABAC Lab along with the intended users for each. A demo video showcasing the main features of ABAC Lab is available at \url{https://youtu.be/KtmSem5jK-A}. 

\begin{table*}[htb]
\centering
\begin{tabular}[b]{|p{2.5cm}|p{7cm}|p{4cm}|}
\hline
\textbf{Feature} & \textbf{Description} & \textbf{Primary Users} \\
\hline
Statistics & Provides dataset-level statistics including number of users, resources, attributes, rules, and permissions. & Researchers / Policy Admins \\
\hline
Log Generator & Generates synthetic access logs with customizable parameters for testing and evaluation. & Researchers \\
\hline
Rule Analysis & Analyzes rule coverage and lists permissions granted by each rule. & Researchers / Policy Admins  \\
\hline
Access Evaluation & Evaluates access decisions for individual or batch access requests based on the policy. & Researchers / Policy Admins  \\
\hline
Data Conversion & Converts policy data into CSV format or serializes it for reuse in experiments. & Researchers \\
\hline
Resource Access & Visualizes the most and least accessed resources to highlight access concentration or restrictions. & Researchers / Policy Admins \\
\hline
Attribute Usage & Analyzes attributes used in rules and generates a heatmap showing permission-attribute relationships. & Researchers / Policy Admins  \\
\hline
\end{tabular}
\smallskip
\caption{Summary of ABAC Lab Features and Intended Users}
\vspace{-2em}
\label{tab:abaclab-features}
\end{table*}

\paragraph{\textbf{Statistics}}

\begin{figure}[htb]
  \centering
\includegraphics[width=0.40\textwidth]{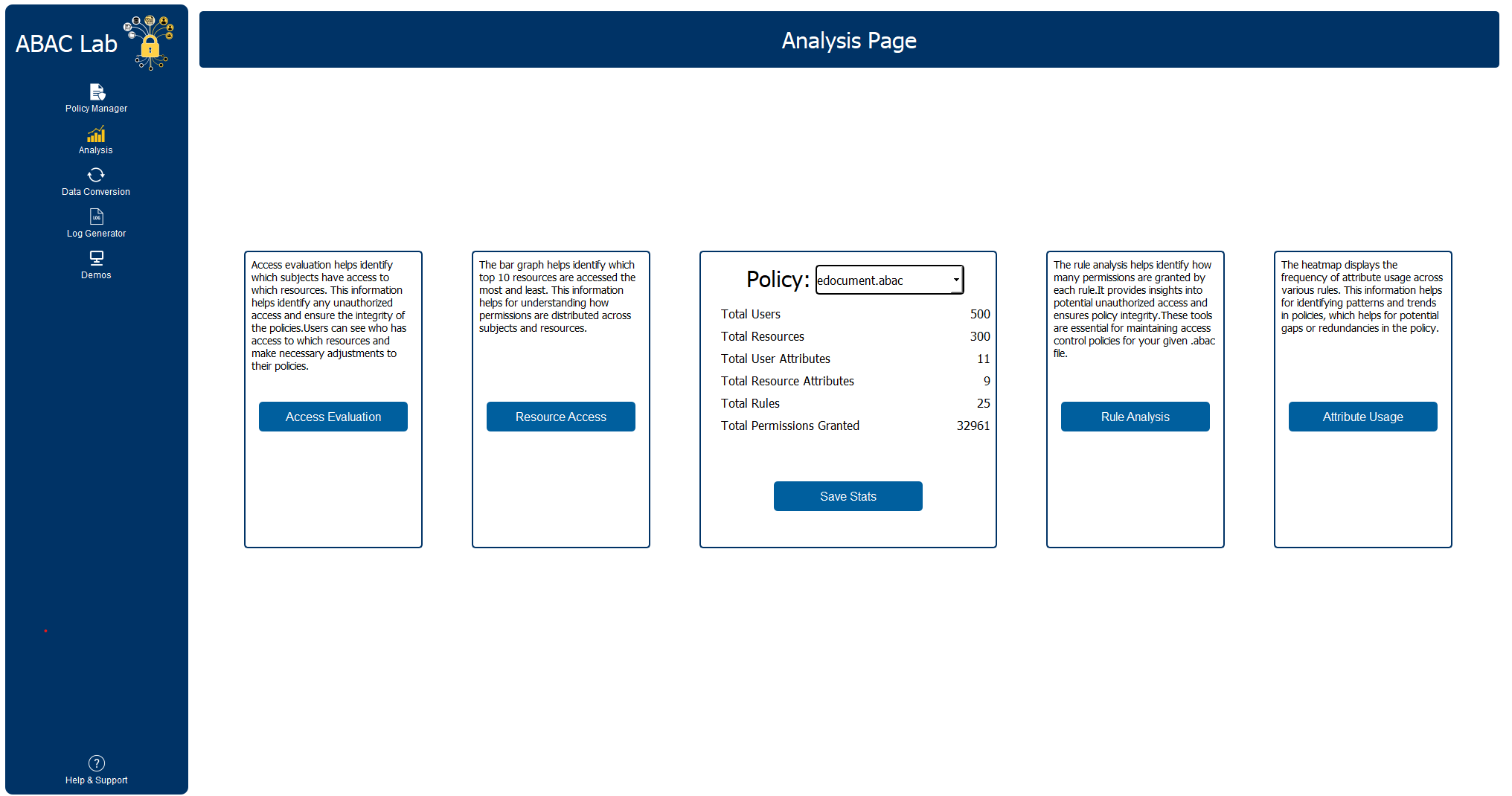}
  \caption{ABAC Lab Statistics and Analysis Home Page}
  \label{fig:abablab-analysis-home}
\end{figure}

The Statistics feature computes key dataset metrics after the dataset is loaded into the system. It reports the number of users, resources, attributes, rules, and permissions, as reported in Table \ref{tab:policy-size}. This feature helps users quickly assess the size and complexity of a dataset, which is particularly useful for evaluating the scalability of policy mining algorithms or reviewing the completeness of ABAC policies. Both researchers and policy administrators benefit from this feature, as it provides a clear overview of the dataset before further analysis or experimentation. Figure~\ref{fig:abablab-analysis-home} shows the Statistics feature and the features home page.

\paragraph{\textbf{Log Generator}}

The Log Generator feature allows researchers to create synthetic access logs based on customizable criteria. Users can specify the total number of log entries and control the percentage of permitted and denied requests. The system generates these logs by evaluating the loaded dataset, using the object model and the input policy rules to determine access decisions. Users, resources, and actions are selected randomly while ensuring the generated logs meet the requested size and permission ratio. As discussed in Section \ref{sec:abac-mining}, this feature supports research on mining policies from access logs by providing controlled, synthetic log data suitable for testing mining algorithms. As discussed in Section \ref{sec:abac-mining}, this feature supports research on mining policies from access logs by providing controlled, synthetic log data suitable for testing mining algorithms. To support broader research scenarios, the feature also allows the injection of noise into the logs in the form of over-permission (incorrectly granted) and under-permission (incorrectly denied) entries. This added flexibility enables controlled experimentation for tasks such as anomaly detection and mining ABAC policies under noisy conditions \cite{buirebacminingnoise}.

\paragraph{\textbf{Rule Analysis}}

The Rule Analysis feature provides detailed evaluation of the policy rules and their coverage. It supports two modes: the first analyzes the rules from the loaded dataset, showing how many permissions are granted by each rule and listing all permissions covered. The second mode both allows users to input a separate set of rules and computes the total number of permissions those rules grant based on the attribute data in the input policy, also listing the specific permissions covered by each rule.

This feature helps evaluate individual rule coverage, ensuring no unintended permissions are granted and all intended permissions are properly captured. It is particularly useful in ABAC policy mining research for assessing the quality of mined rules and comparing them to the ground truth. It also helps researchers and policy administrators validate the correctness and consistency of policies within datasets, ensuring they behave as expected. This validation is especially valuable when developing new datasets for inclusion in the ABAC Lab repository, as it helps verify that generated policies align with intended access control requirements.

\paragraph{\textbf{Access Evaluation}}

The Access Evaluation feature provides users with two options to test access decisions based on the loaded ABAC policy. In Manual Check, users can specify any combination of subject, resource, and action, as each field is optional. This flexibility allows for both targeted and broad access checks based on different analysis needs. The system then evaluates the input against the policy rules and returns the access decision or relevant matching permissions. In File Check, the feature processes a batch of access requests provided in an input file and evaluates each request against the policy. Figure \ref{fig:access-eval-page} in Section \ref{sec:figures} shows the graphical interface for the Access Evaluation feature when evaluating access requests from a file using the university dataset.

This feature allows users to check the permissions of a specific user across all resources or a particular resource. It also supports querying a resource to identify users with access. Policy administrators can use it for auditing, validating policy correctness, and troubleshooting policy configurations. Researchers benefit from this functionality when debugging, testing dataset behavior, and validating the output of mining or analysis algorithms. It also helps identify potential over-permission or under-permission issues.

% \begin{figure*}[tbp]
% \begin{tabular}[t]{@{}l@{}}
%   \centering
% \includegraphics[width=0.7\textwidth]{access-eval-uni.png}
% \end{tabular}
%   \caption{hello}
%   \label{fig:access-eval-gui}
% \end{figure*}

\paragraph{\textbf{Data Conversion}}
The Data Conversion feature provides two options to export ABAC policy datasets into formats suitable for further analysis or integration with other tools. The first option allows users to convert attribute data into CSV files, with attributes stored as columns and corresponding user or resource attribute values are placed in rows, while policy rules are saved separately in a text file. The second option allows users to serialize the parsed attribute data and rules, stored internally as Python dictionaries, into a binary file that can be deserialized and reloaded later.

This feature is particularly useful for researchers who need datasets in a uniform, ready-to-use format for their ABAC policy mining or analysis algorithms. The Python-serializable export aligns with ABAC Lab’s Python implementation and Python’s widespread use in data analysis, making it easier for researchers to integrate exported datasets directly into their workflows. Consistent data format also enables easier dataset reuse for evaluation and comparison, supporting reproducibility and benchmarking across different ABAC policy analysis methods.

\paragraph{\textbf{Resource Access}}
The Resource Access feature visualizes how widely resources are accessible within the loaded policy by generating bar graphs that display the top 10 resources with the highest and top 10 with the lowest number of users granted access. This allows users to easily spot which resources are broadly accessible and which are more restricted.

This feature helps identify critical or sensitive resources that may be overly exposed and require further review, and resources that are highly restricted and potentially underutilized. Policy administrators can use this information for auditing and reviewing resource exposure, supporting risk assessments and ensuring that access is aligned with organizational policies. Researchers benefit from understanding access distribution patterns, which can inform dataset analysis, evaluation of policy mining algorithms, and testing models under varying access scenarios.

\paragraph{\textbf{Attribute Usage}}
The Attribute Usage feature analyzes the policy to examine how attributes contribute to permission assignments. It calculates the number of permissions each rule grants and identifies the specific attributes used in each rule. The results are visualized through a heatmap that shows the relationship between rules and attributes, where the intensity of each cell reflects the number of permissions influenced by that attribute within a rule.

This feature helps users understand which attributes play a significant role in policy decisions and which are less impactful. Policy administrators can use this analysis to review attribute usage for potential policy simplification, ensuring that unnecessary or redundant attributes are not included in critical rules. It also supports auditing efforts by highlighting attributes that heavily influence access decisions. Researchers benefit by gaining insights into attribute importance and dependencies within the dataset, which is valuable for developing or evaluating policy mining algorithms and understanding how different attributes affect policy behavior. This analysis can also inform improvements to dataset design or policy refinement efforts.
% !TeX root = main.tex
\section{Conclusions and Future Work}
\label{sec:conclusion}

This paper introduced ABAC Lab, an open-source platform designed to support ABAC policy mining and analysis by integrating datasets, tools, and new real-world case studies. ABAC Lab addresses key challenges in the field, including the limited availability of publicly accessible ABAC datasets, the lack of standardized formats, and the need for tools that enable systematic policy evaluation. 

By offering a unified environment for ABAC dataset exploration and analysis, ABAC Lab takes an important step toward standardizing resources for ABAC policy mining and analysis research. The platform aims to serve as a common foundation for reproducible experiments and enable researchers to evaluate and compare new methods more effectively. To support ongoing growth, ABAC Lab is designed as a collaborative resource, and we encourage researchers to contribute additional datasets or tools to expand the repository and support diverse ABAC research directions.

In future work, we plan to expand ABAC Lab’s analysis features, including policy comparison, conflict detection, and enhanced visualizations. We also aim to contribute more datasets, including those derived from real-world policies using techniques like automated extraction from natural language descriptions and potential collaborations with industry partners. These expansions will strengthen ABAC Lab as a resource for the community and future research.

\svonly{
\begin{acknowledgements}
\thanksText
\end{acknowledgements}}

%\svonly{\myparagraph{Conflict of Interest} The authors declare that they have no conflict of interest.}

% svjournal's spphys bibstyle does not print titles of conference papers.  I used bibstyle plain instead, since it also uses numeric references.
% 
% \ieeeonly{\bibliographystyle{IEEEtran}}\acmonly{\bibliographystyle{ACM-Reference-Format}}\lncsonly{\bibliographystyle{splncs04}}\articleonly{\bibliographystyle{alpha}}\svonly{\bibliographystyle{plain}}
% \bibliography{references}
%\IfFileExists{references.bib}{\bibliography{references}}{\bibliography{../../references}}
\acmonly{\bibliographystyle{ACM-Reference-Format}}
\bibliography{references}

\appendix
% !TeX root = main.tex
\section{Resource Links And Screenshots}
\label{sec:figures}
This section includes resources that are referenced throughout the paper and an additional screenshot of the ABAC Lab interface as discussed in Section \ref{sec:abac-lab}.

\paragraph{\textbf{ABAC Lab Repository and Demo Video Links }}
%Includes install instructions, dataset files, and analysis tools. The repository also includes detailed .abac format specifications, attribute explanations, and rule structures for each dataset.

\begin{itemize}
    \item \url{https://github.com/ABAC-Lab-Admin/ABAC-Lab}
    \item \url{https://youtu.be/KtmSem5jK-A}
\end{itemize}

% \paragraph{\textbf{Demo video: }}
% %A short video demonstrating the main features of ABAC Lab and how to use the application.

% \begin{itemize}
%     \item \url{https://youtu.be/KtmSem5jK-A}
% \end{itemize}

% \section{Interface Screenshots}
% \label{sec:figures}
% This section include some screenshots of the ABAC Lab interface as discussed in Section \ref{sec:abac-lab}.

\begin{figure}[htb]
  \centering
\includegraphics[width=0.47\textwidth]{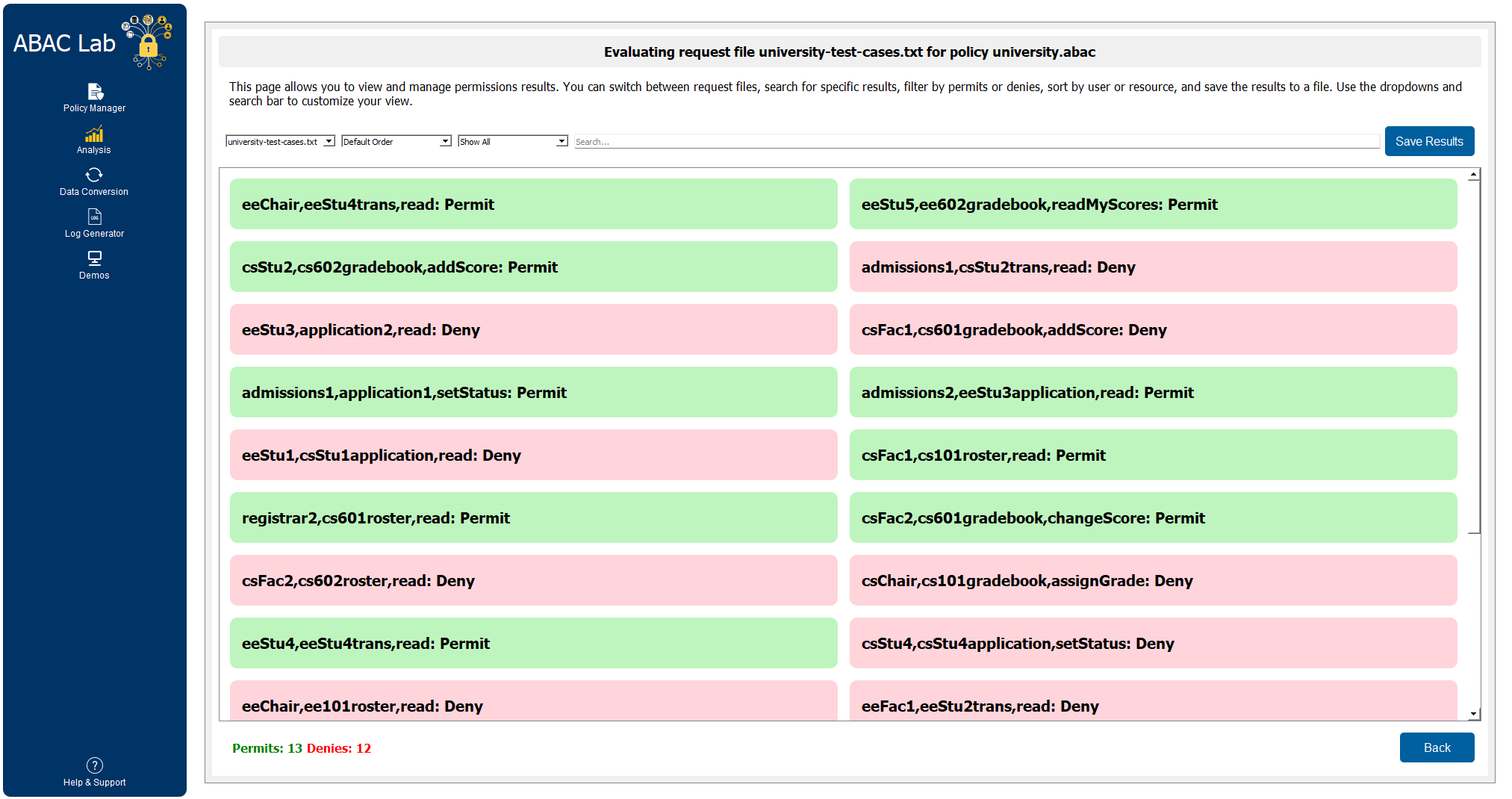}
  \caption{ABAC Lab Access Evaluation Page}
  \label{fig:access-eval-page}
\end{figure}

 \end{document}